# Fully CMOS-compatible passive TiO$_2$-based memristor crossbars for in-memory computing

Abdelouadoud El Mesoudy[1,2], Gwénaëlle Lamri[1,2], Raphaël Dawant[1,2], Javier Arias-Zapata[1,2], Pierre Gliech [1,2], Yann Beilliard[1,2]*, Serge Ecoffey[1,2], Andreas Ruediger[3], Fabien Alibart[1,2,4], Dominique Drouin[1,2]

[1] Institut Interdisciplinaire d'Innovation Technologique (3IT), Université de Sherbrooke, 3000, boulevard de l'Université, Sherbrooke (Québec) J1K OA5, Canada

[2] Laboratoire Nanotechnologies Nanosystèmes (LN2) – CNRS, Université de Sherbrooke – 3000, boulevard de l'Université, Sherbrooke (Québec) J1K 0A5, Canada

[3] Institut National de la Recherche Scientifique (INRS), centre Énergie, Matériaux, Télécommunications, 1650 Boulevard Lionel-Boulet, Varennes (Québec) J3X 1S2, Canada

[4] Institute of Electronics, Microelectronics and Nanotechnology (IEMN), Université de Lille, 59650 Villeneuve d'Ascq, France

*Corresponding author: yann.beilliard@usherbrooke.ca

**Abstract**

Brain-inspired computing and neuromorphic hardware are promising approaches that offer great potential to overcome limitations faced by current computing paradigms based on traditional von-Neumann architecture. In this regard, interest in developing memristor crossbar arrays has increased due to their ability to natively perform in-memory computing and fundamental synaptic operations required for neural network implementation. For optimal efficiency, crossbar-based circuits need to be compatible with fabrication processes and materials of industrial CMOS technologies. Herein, we report a complete CMOS-compatible fabrication process of TiO$_2$-based passive memristor crossbars with 700 nm wide electrodes. We show successful bottom electrode fabrication by a damascene process, resulting in an optimised topography and a surface roughness as low as 1.1 nm. DC sweeps and voltage pulse programming yield statistical results related to synaptic-like multilevel switching. Both cycle-to-cycle and device-to-device variability are investigated. Analogue programming of the conductance using sequences of 200 ns voltage pulses suggest that the fabricated memories have a multilevel capacity of at least 3 bits due to the cycle-to-cycle reproducibility.

# Introduction

Artificial intelligence (AI) based on second-generation artificial neural networks (ANNs) has become ubiquitous thanks to the ever-increasing computational power, availability of datasets, and breakthroughs in learning methods. Deep learning (DL) [1] enabled by deep neural networks (DNNs) has impacted a wide range of sectors, including computer vision, text translation, healthcare, finance, and automotives. Importantly, the operation of DNNs relies heavily on vector-matrix multiplications (VMMs). Executing such fundamentally parallel operations with current von-Neumann computing platforms [2] causes significant performance and energy consumption issues, mainly induced by the serial data transmission between memory and logic cores. Large-scale and energy-inefficient cloud computing solutions are thus mandatory for DNN training, which represents a technological roadblock, hindering the development of AI-based portable and edge computing applications. Efforts are thus focused towards the development of emerging non-volatile memories (NVMs), especially those based on resistance switching (also called memristors) [3], and an in-memory computing paradigm [4] for energy-efficient ANNs based on VMM accelerators [5].

The memristor is considered as the fourth fundamental two-terminal circuit element, whose internal conductance state depends on the history of current that flows through it [6]. Several memristor classification exist depending on the resistive switching mechanism. Among them, redox-based memristive systems have attracted much attention due to their superior performance in terms of energy consumption and switching speed [7]. These memristor devices rely on ionic and thermal mechanisms involving the formation/rupture of conductive filaments with high concentration of oxygen vacancies [8]. In the case of $TiO_2$-based systems, it has been demonstrated that these filaments are composed of $Ti_4O_7$ Magnéli phase [9] that exhibits a metallic behaviour near room temperature [10].

Furthermore, when organized in crossbar array architecture, memristor devices can naturally perform VMM operations following Ohm's law for multiplication and Kirchhoff's law for current summation. In fact, crossbar arrays of memristors used to both store the synaptic weights of the ANN and perform VMM are very attractive, and simulations of such an approach indicate that several orders of magnitude higher speed-energy efficiencies could be achieved with memristor-based ANNs [11].

For these novel brain-inspired circuits to be scalable, materials and fabrication processes of memory devices must be compatible with the back-end-of-line (BEOL) of industrial CMOS technologies. In that scope, resistive memories based on transition metal oxides are one of the most promising devices owing to their synaptic-like gradual switching behaviour [12]. Recent demonstrations of hardware-implemented neural networks involving this type of memory include either monolithically co-integrated CMOS-memristor circuits [13] or passive memristor crossbar arrays on silicon [14]. However, most of the fabrication processes reported in the literature rely on lift-off techniques and/or metals such as Pt, Pd or Au. Such approaches suffer from several limitations: (i) these techniques are incompatible with industry standards because of material properties and the yield of the process; (ii) Crossbar arrays are very sensitive to electrode resistance and thus benefit from high aspect ratio metal lines that require advanced patterning, etching and filling processes. (iii) Electrode topographies from lift-off techniques present edge roughness and defects, leading to non-uniform electric fields that can increase the variability in switching.

When studies propose crossbar manufacturing approaches that seem to be compatible with BEOL integration, precise metrology details that allow for the evaluation of the overall fabrication process are often missing. Kim et al. [15] presented a complete work for the fabrication and characterisation of BEOL-compatible passive memristor crossbars. Their bottom electrode fabrication process includes 6 different steps: Ti/Al/TiN deposition, lithography, etching, $SiO_2$ deposition, and chemical mechanical polishing (CMP) combined with an etch-back process. However, the evaluation of the fabrication process has only been based on scanning electron microscopy (SEM) images, thus providing few details on the planarisation level or the electrode surface and edge roughness.

In this study, we present a complete fabrication process of $TiO_{2-x}$ based memristor crosspoints and up to 8×8 passive crossbars arrays. For bottom electrode fabrication, we used a BEOL-compatible *nanodamascene* process [16], which involves 4 process steps: lithography, plasma etching, TiN deposition, and CMP. Furthermore, morphological characterisation combining atomic force microscopy (AFM) and SEM were used to evaluate the electrode quality in terms of roughness and planarisation. The electrical performance of the devices was evaluated using quasi-DC sweeps and voltage pulse measurements, showing synaptic behaviour emulation. In this study, the electrical characterisation part was conducted on single crosspoints as the crossbar fabrication constituted mainly a process demonstration.

**Methods**

Fig. 1 shows the fabrication process of $TiO_2$-based memristor crosspoints and crossbars, which was conducted on a 22×22 mm² silicon sample covered with a 600 nm thick SiN layer deposited by plasma-enhanced chemical vapor deposition (PECVD). The 700 nm wide (1.4 µm pitch) TiN bottom electrodes were fabricated using four different steps. The lithography step was performed using a RAITH150-Two e-beam writer. The bottom electrode patterns were transferred to the SiN layer by plasma etching using $CF_4$, $H_2$, and He chemistry with a flow of 140/12/14 sccm and a power of 100/50 W for coil and platen, respectively [17] (Fig. 1a). The etching time was calibrated to obtain 400 nm deep SiN trenches. The latter were filled with 600 nm thick TiN deposited by reactive sputtering, exhibiting a resistivity of 160 µΩ.cm (Fig. 1b). CMP was used to remove the excess TiN and to planarise the samples, resulting in embedded TiN electrodes in SiN, as shown in Fig. 1c. To reduce leakage current in the devices, 1.4 nm of $Al_2O_3$ was deposited by plasma-enhanced atomic layer deposition with a Picosun R-200 advanced ALD system. Trimethylaluminium (TMA) was used as a precursor. A 30 nm thick sub-stoichiometric $TiO_{2-x}$ switching layer was deposited using a Plasmonique SPT 320 sputtering tool. Direct current reactive sputtering of a Ti target was performed in an $Ar/O_2$ gas mixture using flow rates of 40 and 2 sccm, respectively. The working pressure and sputtering power were set to 6.8 mTorr and 150 W, respectively. The top electrode stack composed of a Ti/TiN bi-layer (10 nm/30 nm) was then deposited in the same sputtering chamber. The Ti layer was used in order to create an oxygen vacancy reservoir and an ohmic interface with $TiO_2$ layer, while the TiN layer acts as an inert electrode and an oxygen diffusion barrier. To further decrease the TE resistance, a 400 nm thick Al layer was deposited by e-beam evaporation (Edwards Auto 306), as shown in Fig. 1d. The 700 nm wide tope electrodes were defined by e-beam lithography using the same parameters as for the bottom electrodes. Both the top electrode stack and the switching layer were patterned by plasma etching using $BCl_3/Cl_2/Ar$ chemistry with a 10/10/10 sccm flow rates and 500/50 W for coil/platen power, respectively (Fig. 1e). It is worthwhile noticing that e-beam lithography steps can be routinely performed by standard deep UV lithography stepper tools in industrial foundries.

Single memristor crosspoints, as well as crossbar arrays, were successfully fabricated using our process (Fig. S1). Physical–chemical analysis was performed using SEM inspections and energy dispersive X-ray spectroscopy (EDX), revealing well-defined electrode for 8×8 crossbar arrays (Fig. 1f). A cross-section SEM investigation of one of the memristor devices was performed using a focused ion beam tool. Fig. 1g shows an SEM cross-section of the device composed of the $Al_2O_3/TiO_{2-x}$ active layer sandwiched by a plasma-etched top electrode and a planar bottom electrode embedded into the SiN layer, resulting from the damascene process. The TiN of the bottom electrode seems porous at the edge of the trench. While this did not cause problems in our study, it may require further attention.

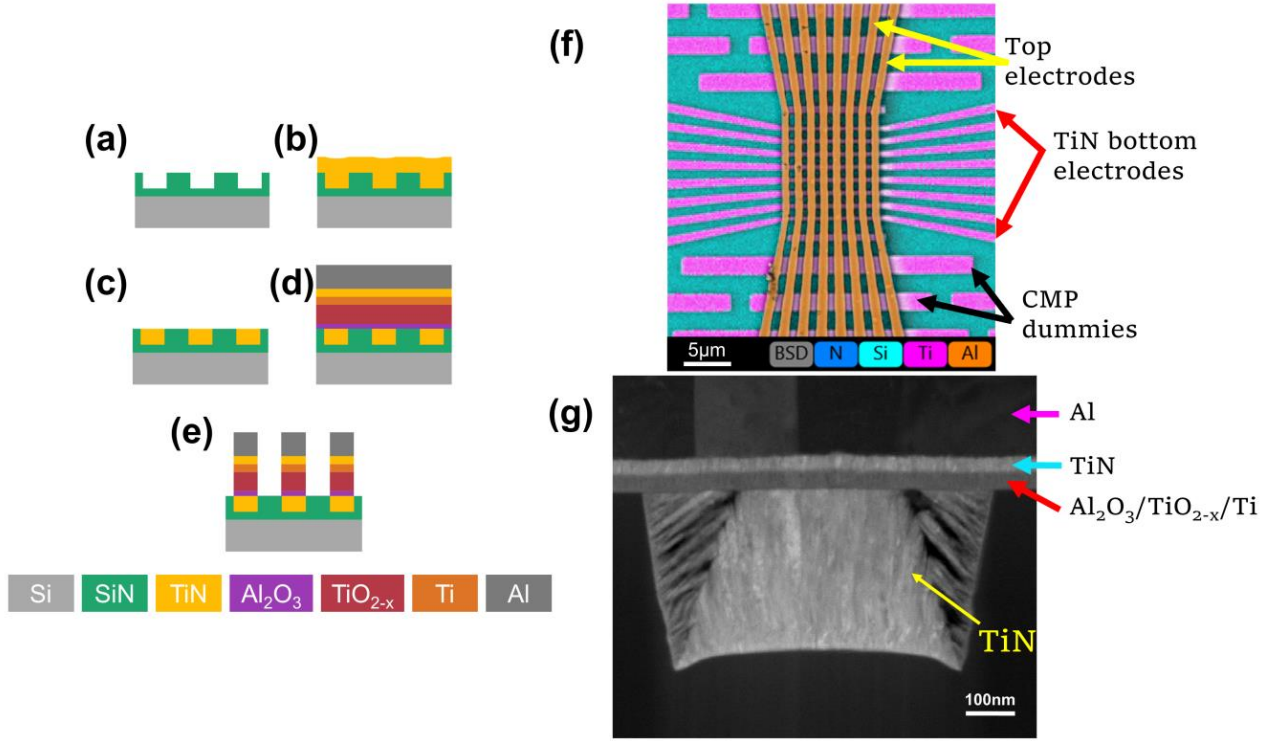

**Fig. 1.** Process flow and morphological characterisations of the scalable passive CMOS-compatible TiO$_2$-based memristor crossbars. **(a–c)** Damascene process for bottom electrode patterning, including **(a)** SiN deposition, e-beam lithography, plasma etching, **(b)** TiN deposition, and **(c)** CMP. **(d)** Switching layer (Al$_2$O$_3$/TiO$_{2-x}$) and top electrode stack (Ti/TiN/Al) deposition. **(e)** Top electrode patterning performed by e-beam lithography and plasma etching. **(f)** SEM-EDX top-view of an 8×8 crossbar array. (g) SEM cross-section of a memristor.

## Results and Discussion

Device-to-device (D2D) and cycle-to-cycle (C2C) variability are considered as one of the main challenges hindering the widespread use of memristor-based circuits in a wide range of in-memory computing applications [4][18]. Interestingly, depending on the neural network training strategy, memristor variability may be tolerated or even exploited for some applications [19]. In any case, an on-chip (*in situ*) training strategy is tolerating memristor variability as it has the advantage of using the full range of memristor conductance, and would allow the learning algorithm to account for device variations and non-working devices. On the contrary, an off-chip (*ex situ*) training strategy is more sensitive to device variations since the training is performed using a software-based neural network where trained synaptic weights need to be transferred to the network [20]. In this case, the off-chip training accuracy can be largely affected if there are non-working devices or a drift in device conductance programming [21].

Some of these device variations can be readily attributed to imperfections in the fabrication process [22][23]. To address this issue, several approaches have been proposed at the circuit level, thus leading to more complex circuitry [24]. The variability limitation can also be addressed by optimising the fabrication processes since the memristor's electrical performance can be largely affected by the quality of the electrodes and the interfaces. Mainly, the electrode surface and edge roughness have proven to be a key source of the memristor's electrical degradation [25], probably due to electric field enhancement at the surface asperities, especially during the electroforming step [26]. To improve the fabrication quality of the electrodes, and thus their uniformity, we developed a damascene process based on a highly controlled CMP step to planarise a 700 nm wide TiN BE. Fig. 2a shows an SEM image of 8 bottom electrodes after CMP. The AFM scan in Fig. 2b shows a planarised TiN-SiN structure. Because of its lower removal rate, TiN electrodes are slightly above the SiN surface by an average of 8 nm, as demonstrated by the extracted profile in Fig. 2c. The measured root mean square (RMS) electrode surface roughness is as low as 1.1 nm. The damascene process enables the fabrication of high-quality and reproducible bottom electrodes embedded in SiN, with low surface roughness and dishing. However, further optimisations are still needed to reduce the electrode edge effect as it may be considered another source of variability. The developed fabrication process can easily be adopted by silicon foundries for memristor integration in the BEOL of CMOS chips.

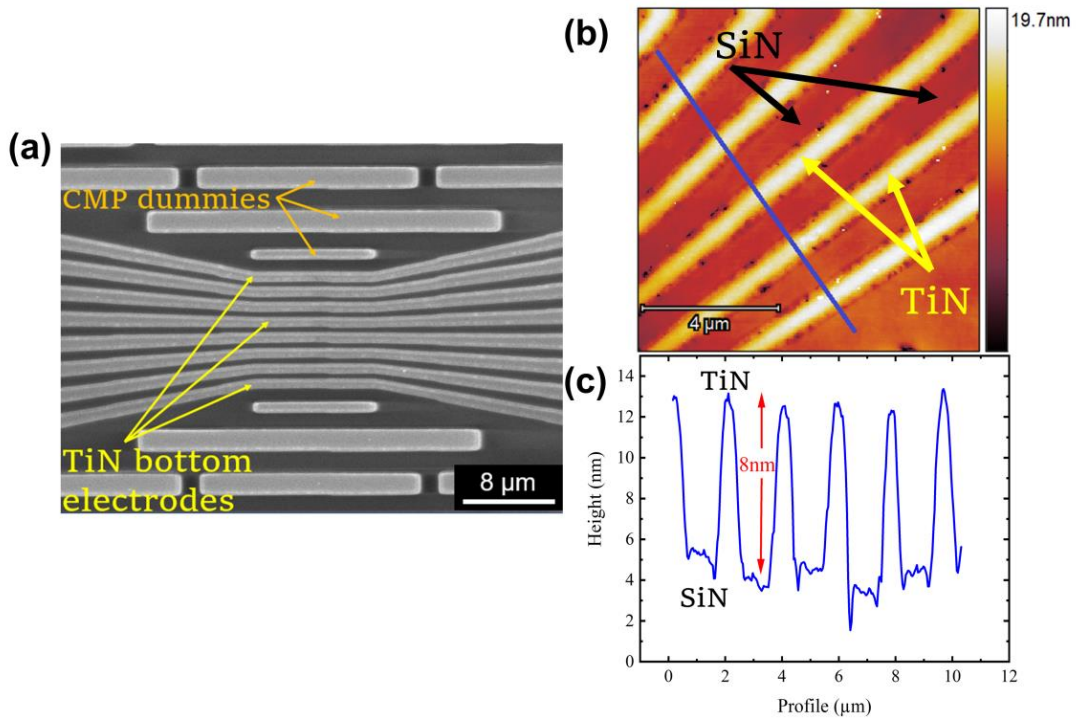

**Fig. 2.** Morphological characterisations of the TiN bottom electrodes fabricated by the damascene process. **(a)** SEM top-view of the planarised TiN bottom electrodes using CMP. **(b)** AFM 10×10 µm² scan of the embedded TiN bottom electrodes. The blue line shows the position of the extracted profile. **(c)** Extracted profile illustrating the SiN-TiN planarisation level.

To evaluate the electrical behaviour of the fabricated devices, electrical characterisations were conducted on 10 single crosspoints with 700 nm wide electrodes (Fig. S1a). Quasi-DC and pulsed measurements were performed using a Wentworth MP900 probe station and a Keithley SCS4200 parametric analyser. For all devices, the bottom electrodes were grounded, and the signals were applied to the top electrodes. Prior to any resistive switching, a voltage-controlled electroforming step was conducted with a current compliance of $I_c = 300$ µA. The current limitation was carried out using an external off-the-shelf transistor in series with the tested devices.

The typical bipolar resistive switching behaviour of our TiN/Al$_2$O$_3$/TiO$_{2-x}$/Ti/TiN/Al memristor devices is shown in Fig. 3a, where 30 quasi-DC switching cycles were performed. In this case, the full switching transition was obtained by applying a negative quasi-DC voltage sweep for the RESET transition (between 0 and −1.4 V), followed by a positive quasi-DC current sweep for the SET transition (between 0 and 1.3 mA). Note that the SET and RESET switching voltages ($V_{SET}$ and $V_{RESET}$ respectively), defined by the inflexion point of d$I$/d$V$, show a low standard deviation over 30 cycles as the average values are 0.95 ± 0.048 V ($V_{SET}$) and −0.84 ± 0.052 V ($V_{RESET}$). This result suggests a low C2C variability (see inset). Furthermore, the ratio between the high resistive state (HRS) and the low resistive state (LRS) remains slightly higher than 10 after 30 cycles, as shown in Fig. 3b. It is worthwhile noticing that quasi-DC sweeping protocol is not adapted for testing device endurance, as electrical stress is applied to the memristor for longer periods. To evaluate device endurance, voltage pulses were used resulting in 10,000 switching cycles while maintaining satisfying resistance window — see Supplementary Fig. S2.

The D2D variability was investigated using 10 crosspoints fabricated on the same sample as large device variability can be a major limitation towards robust ANN implementation. Fig. 3c and Fig. 3d demonstrate the average values of the switching voltages ($V_{SET}$, $V_{RESET}$) and resistance states ($R_{ON}$, $R_{OFF}$), respectively, for 10 different crosspoints. The error bars indicate their standard deviations for 10 cycles.

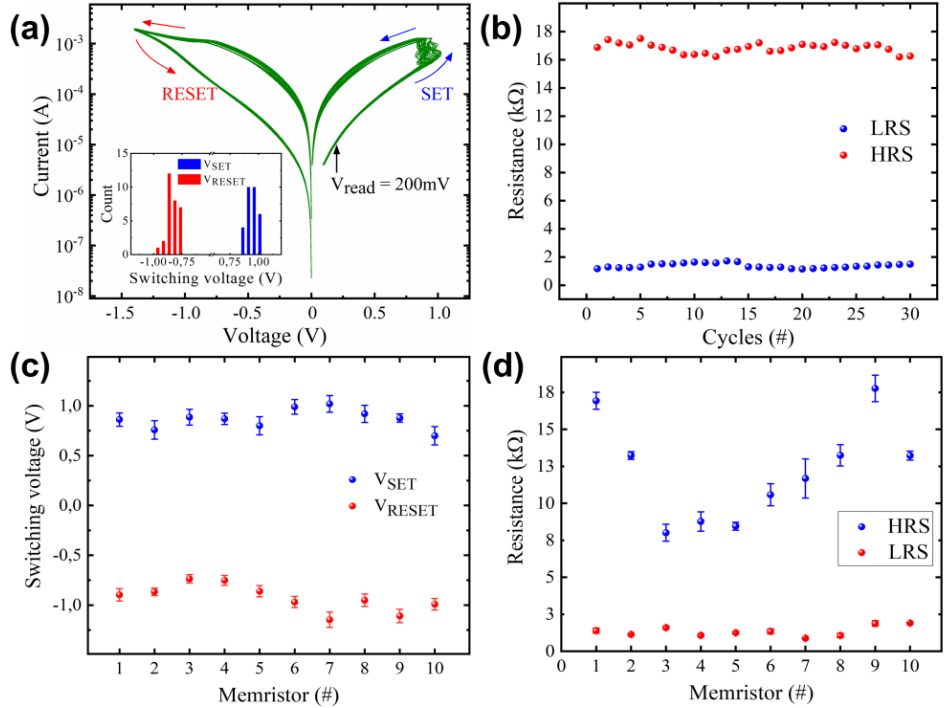

**Fig. 3.** Quasi-DC characterisations and statistical analysis of memristor crosspoints. **(a)** *I–V* sweeps showing typical bipolar resistive switching. The inset shows a histogram of the $V_{SET}$ and $V_{RESET}$ distribution for 30 cycles. **(b)** Resistance ratio ($R_{ON}/R_{OFF}$) evolution over 30 cycles. D2D variability of $V_{SET}$, $V_{RESET}$ **(c)** and HRS, LRS **(d)** for 10 different memristors. The symbols represent the average values, and the error bars indicate their standard deviations for 10 cycles.

The switching voltage average values are 0.86 ± 0.16 V and −0.92 ± 0.22 V for $V_{SET}$ and $V_{RESET}$, respectively, demonstrating low variability and operating voltages compatible with CMOS integration. The switching voltage distribution is sufficiently narrow to avoid disturbing the half-selected devices in a crossbar implementation. Furthermore, batch-to-batch variability data (Fig. S3) shows slight deviations in operating voltages over different batches, which might be attributed to variations in the overall fabrication process. It is worthwhile noticing that the 4 different fabricated batches present a production yield of 75 %, 75 %, 71 % and 70 % using an academic clean room facility. Further optimization of the fabrication process is thus needed in order to improve the memristor yield for practical applications. Especially, limiting the amount of uncontrolled defects introduced during the fabrication process steps, including film deposition and patterning steps, should greatly enhance the reliability and yield of the memristors.

On the other hand, the resistance states show a relatively large D2D variability, especially for $R_{OFF}$ (Fig. 3d). In fact, the HRS seems to vary between 8 kΩ and 18 kΩ, which can be attributed to the variability of our forming process despite the use of an external off-the-shelf transistor. Small variations or overshoots of the current compliance induced by capacitive effects in the transistor could indeed lead to the formation of conductive filaments with different geometries from one device to the other. Contrary to the LRS, the HRS is more sensitive to this geometry variation [27], which results in D2D variability. Nevertheless, even the lowest resistance range of 1.5 kΩ to 8 kΩ (ratio ~5) is enough to encode the synaptic weights of an ANN for classification tasks [28]. Different strategies can be used to increase the device resistance range, such as: (i) reducing the leakage current of the switching layer matrix as compared the HRS current will result in a higher HRS state. This can be done by optimizing the switching layer oxygen deficiency [11], or by increasing $Al_2O_3$ layer thickness. However, both strategies can influence the electroforming voltages of our devices. An optimal combination of $TiO_2$ oxygen deficiency and $Al_2O_3$ layer thickness needs to be chosen. On the other hand, (ii) reducing the current overshoots during the forming process could also result in an increase of the filament resistance, thus improving the resistance window. Furthermore, (iii) the overall resistance range of our devices could be increased by scaling down the electrode dimensions. This can be easily implemented due to the scalability of our process, which would further limit the leakage currents. However, scaling down electrode's dimensions will result in an increase of line resistance which should be adjusted by increasing the electrods's aspect ratio. Shuang et al. demonstrated a memristor crossbar array with 2 nm feature size showing extreme scalability of memristors technology, exhibiting low operational current and a large resistance range [29]. This suggests that scaling down the feature size of our devices could potentially lead to a considerable increase in resistance window and reduce the power consumption by lowering the operating current.

Besides binary bipolar switching, we investigated if conductance of our memristors could be tuned gradually to efficiently emulate the weight modulation of synapses in ANNs [30]. To achieve analogue operations, gradual SET (RESET) transitions were first performed by applying multiple quasi-DC current (voltage) sweeps of increasing amplitude, as shown in Fig. 4a. The incremental SET transitions were obtained by gradually increasing the current sweep amplitude from 0 to 1.3 mA with a step of 100 µA. Likewise, gradual RESET was obtained using voltage sweeps of increasing amplitude from 0 to −1.4 V with a step of 100 mV. The programmed multilevel conductance states are non-volatile and can be retained for more than $10^4$ s, as shown in Fig. S4.

Conductance tuning was also performed on our devices using short voltage pulses. This type of operation provides a better way to reduce the influence of thermal effects and matches the operating conditions of integrated devices. The results of the pulse-based memristor programming are regrouped in Fig. 4b and Fig. 4c, where the long-term depression (LTD) and long-term potentiation (LTP), respectively, can be observed for different voltage values. The pulse sequence was comprised of a 200 ns write pulse followed by a 1 ms read pulse. In total, 400 write pulses were applied to the device for each voltage amplitude. After each pulse sequence, a quasi-DC SET (RESET) sweep was performed to regain the initial high (low) conductance state. For each voltage value, this whole protocol was executed five times in order to evaluate the C2C variability (error bars in Fig. 4b,c).

For the RESET operation (Fig. 4b), the conductance depression varies depending on the applied negative voltage. Low voltage amplitudes (−1.3 V, −1.4 V) result in a relatively small and linear change in the conductance state. Such behaviour is highly desired for online tuning of the synaptic weight, which was proven to improve the learning accuracy of neural networks [31]. Increasing the voltage amplitude (−1.5 V to −1.7 V) results in a more important depression dynamic at the cost of a less linear weight change. A further increase in voltage amplitude (−1.8 V, −1.9 V) results in an abrupt conductance change, which slows down after the first few pulses and then saturates afterwards.

Fig. 4c represents the conductance potentiation using the same pulse sequence but with positive voltage. No conductance change was observed below 1.1 V. However, conductance potentiation appears to be more abrupt compared to depression dynamics. This may be attributed to the electric-field-driven migration of oxygen vacancies assisted by Joule heating and resulting in a positive feedback effect, leading to an abrupt conductance change following the first few pulses [32]. The associated non-linearity could be lowered by optimising the pulse programming scheme [33] or the pulse shape. It has indeed been reported that a slower pulse rising time can induce gradual filament formation, thus leading to a less abrupt conductance change [34][35].

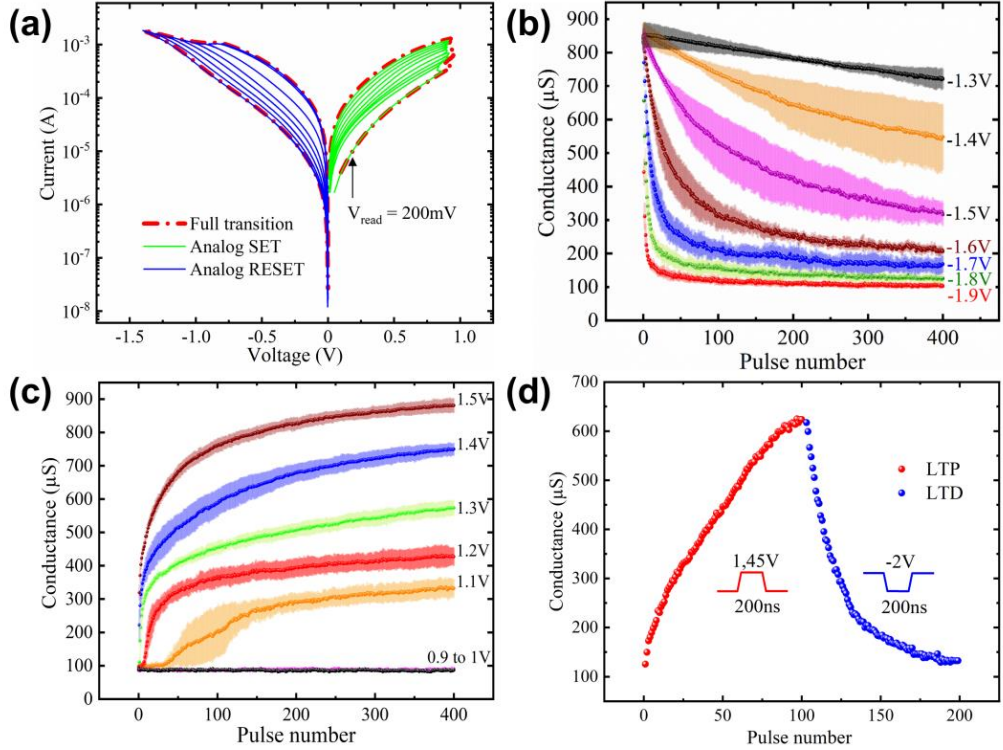

**Fig. 4.** **(a)** Analogue resistive switching obtained by incremental quasi-DC sweeps. In total, 9 gradual SET and 7 gradual RESET increments have been performed. Incremental long-term conductance **(b)** depression and **(c)** potentiation showing the switching dynamics induced by 200 ns long voltage pulses for different voltage values. The symbols and coloured error bars represent the average conductance values and their standard deviations for 5 cycles, respectively. **(d)** Quasi-linear long-term potentiation and depression curves obtained with 200 ns long pulses. The applied voltage amplitudes were 1.45 V and −2 V for the LTP and LTD, respectively.

As described earlier, we evaluated the C2C variability by repeating the LTP and LTD operations 5 times for a given voltage. For each voltage amplitude, the coloured surface represents the standard deviation for five different cycles. Even though a certain amount of C2C variability is noticeable, note that there is no overlap in the programmed conductance range for different voltage amplitudes. This implies that each voltage amplitude can be used to target a specific conductance range. These results along with the quasi-DC sweeps allow to estimate that our devices have a multilevel capacity of at least 3 bits. Finally, Fig. 4d illustrates the continuous potentiation and depression characteristics obtained using 200 pulses in total. The applied voltage amplitudes were 1.45 V and −2 V for potentiation and depression, respectively. Note that the LTP and LTD characteristics are quasi-linear for the range of conductances going from 250 µS to 650 µS. Cycle-to-cycle variability for LTP/LTD characteristic are shown in Supplementary Fig. S5.

As previously stated, a linear conductance change is highly desired for precise modulation of device's synaptic weights. All these results suggest that our memristor devices fabricated using a CMOS-compatible approach based on a damascene process are highly competitive for the implementation of scalable artificial synapses.

**Conclusion**

We have reported a BEOL-compatible fabrication process for TiO$_{2-x}$ based memristors. Crosspoints and crossbar arrays have been fabricated. Successful bottom electrode fabrication by a damascene process has been demonstrated, achieving low surface roughness of 1.1 nm, which allowed for satisfactory resistive switching performance. Analogue conductance switching, cycle-to-cycle and device-to-device variability have been investigated using quasi-DC sweeps and voltage pulse protocols. We have shown that device conductance can be programmed using a sequence of 200 ns pulses of less than 2 V, and can also be controlled to achieve quasi-linear LTP and LTD. Such analogue conductance programming suggests that on-chip learning could be performed with these memristor devices. A multilevel capacity of at least 3 bits could also be used thanks to reproducible C2C characteristics. This work paves the way for CMOS-memristor monolithic 3D integration at an industrial scale, which can offer novel opportunities for the hardware implementation of in-memory computing.

**Acknowledgements**

This work was supported by the Natural Sciences and Engineering Research Council of Canada (NSERC) HIDATA project 506289-2017 and ERC-CoG IONOS (# GA 773228). A.R. gratefully acknowledges financial support through an NSERC discovery grant (RGPIN-2019-07023). This work was also supported by the CHIST-ERA UNICO project and Fond de Recherche du Québec Nature et Technologies (FRQNT). We would like to acknowledge Abdelatif Jaouad, Julien Pezard, and 3IT.Nano platform for their valuable support with device fabrication. We would also like to thank Yuanyang Guo for her assistance with electrical characterisation.

**Declaration of Competing Interest**

The authors declare no conflict of interest.

# Supplementary materials

# Fully CMOS-compatible passive TiO$_2$-based memristor crossbars for in-memory computing

Abdelouadoud El Mesoudy[1,2], Gwénaëlle Lamri[1,2], Raphaël Dawant[1,2], Javier Arias-Zapata[1,2], Pierre Gliech[1,2], Yann Beilliard[1,2]*, Serge Ecoffey[1,2], Andreas Ruediger[3], Fabien Alibart[1,2,4], Dominique Drouin[1,2]

[1] Institut Interdisciplinaire d'Innovation Technologique (3IT), Université de Sherbrooke, 3000, boulevard de l'Université, Sherbrooke (Québec) J1K OA5, Canada

[2] Laboratoire Nanotechnologies Nanosystèmes (LN2) – CNRS, Université de Sherbrooke – 3000, boulevard de l'Université, Sherbrooke (Québec) J1K 0A5, Canada

[3] Institut National de la Recherche Scientifique (INRS), centre Énergie, Matériaux, Télécommunications, 1650 Boulevard Lionel-Boulet, Varennes (Québec) J3X 1S2, Canada

[4] Institute of Electronics, Microelectronics and Nanotechnology (IEMN), Université de Lille, 59650 Villeneuve d'Ascq, France

1- **SEM top-view images of the fabricated devices**

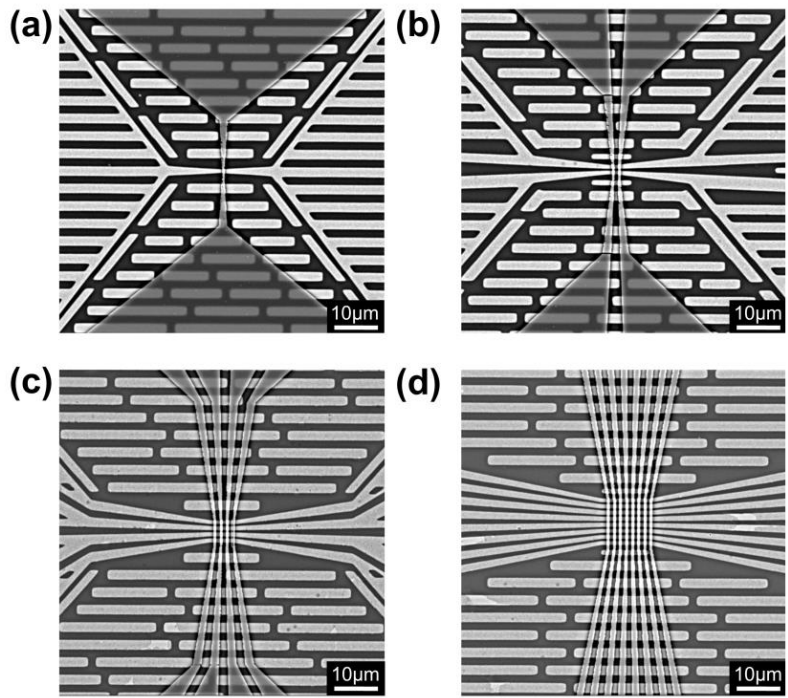

**Fig. S1.** SEM images of the different fabricated memristor devices **(a)** Single crosspoint **(b)** 2×2 **(c)** 4×4 and **(d)** 8×8 crossbar arrays.

## 2- Endurance

To evaluate the endurance of the fabricated devices. Short voltage pulses of 10 µs-width and -2.5 V amplitude for the RESET operation. While for the SET operation, pulses of 1 µs-width and 2 V amplitudes were used.

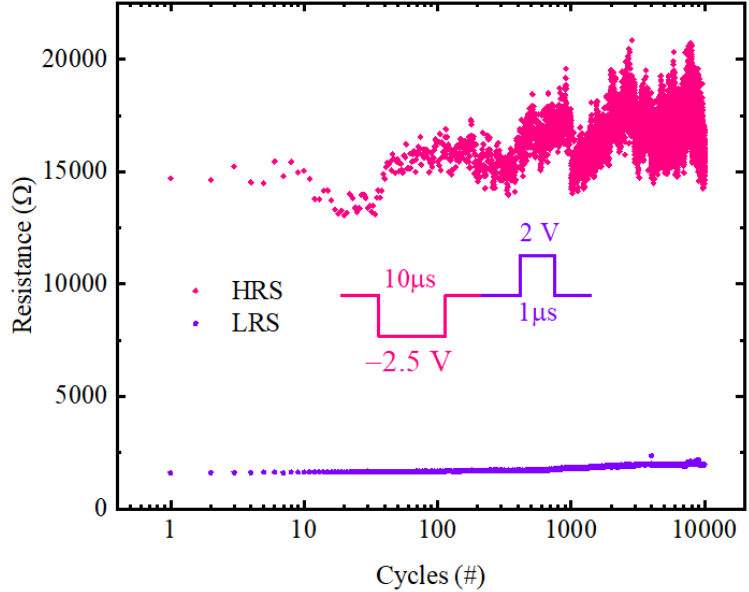

**Fig. S2.** Endurance test using short pulses resulting in 10,000 cycles. For the SET operation, 1 µs-width pulses were used with 2 V voltage amplitude. While for the RESET operation, 10 µs-wide pulses were used with -2.5 V voltage amplitude.

## 3- Batch-to-batch variability

In order to evaluate the batch-to-batch variability, 4 different batches have been fabricated using an academic clean room facility, presenting a production yield of 75 %, 75 %, 71 % and 70 %. Further optimization of the fabrication process is thus needed in order to improve the memristor yield for practical applications. Especially, limiting the amount of uncontrolled defects introduced during the fabrication process steps, including film deposition and patterning steps, should greatly enhance the reliability and yield of the memristors.

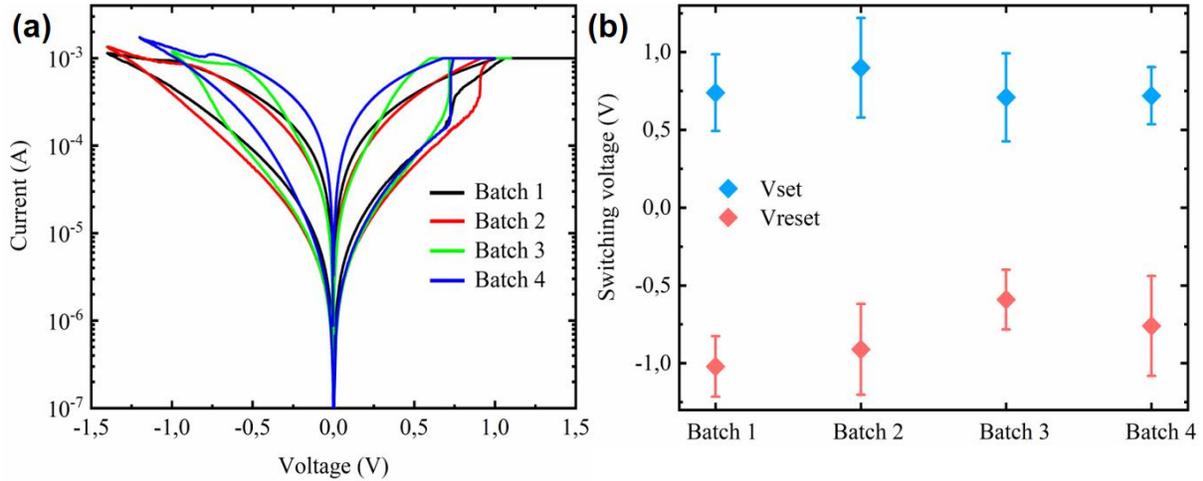

**Fig. S3.** Batch-to-batch variability data extracted from 4 different batches: (a) Quasi-DC bipolar switching characteristics for memrsitor devices fabricated on different batches. (b) Switching voltage values extracted from different batches. A slight deviation can be noticed in switching voltages, which might be due to fabrication process variability. The symbols represent the average switching voltage values and the error bars represent their standard deviation for 10 memristors.

4- **State retention measurements**

Retention measurements were performed at room temperature for more than 8 hours for 8 different resistance states, which demonstrates a multilevel capacity of at least 3 bits. After programming each state, a read pulse (1ms-long) has been applied every 5 minutes in order to sense device's resistance. The programmed states are non-volatile for the entire test duration ($3.10^4$s).

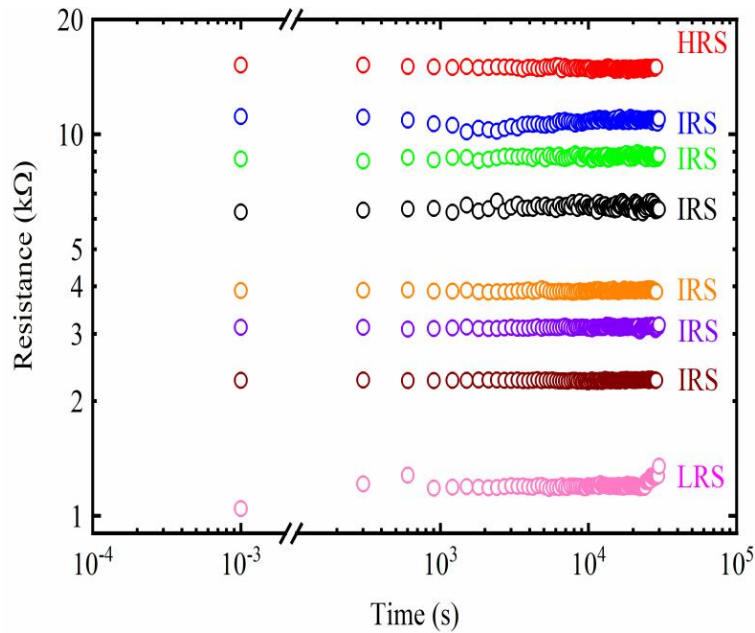

**Fig. S4.** State retention measurements performed on a TiN/Al$_2$O$_3$/TiO$_{2-x}$/Ti/TiN/Al memristor crosspoint. In total, 8 resistance states were tested including HRS, LRS and 6 different intermediate resistance states (IRS).

### 5- Cycle-to-cycle data for long term potentiation and depression

Fig.S5 presents cycle-to-cycle data for the LTP/LTD characteristics. For these tests, 200 ns voltage pulses were used, with voltage amplitudes of 1.4 V and -1.9 V for the LTP and LTD, respectively. One can note that the device switches between 2 stable conductance states of 52 µS and 350 µS. Gradual LTP characteristics is obtained while the LTD presents a sharp transition. As discussed earlier, the pulse scheme needs to be optimized in order to obtain more gradual characteristics.

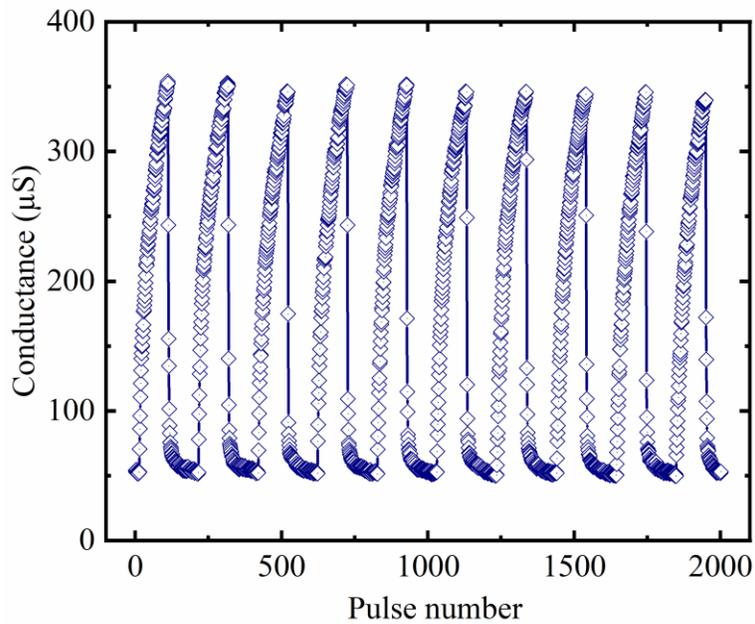

**Fig. S5.** Cycle-to-cycle data for the LTP/LTD characteristics using 200ns voltage pulses. Voltage amplitudes of 1.4 V and -1.9 V were used for conductance potentiation and depression, respectively.